# ZnO Nanowire Arrays Decorated with Titanium Nitride Nanoparticles as Surface Enhanced Raman Scattering Substrates

Y. Rajesh,[1] M.S.S. Bharati,[2] S. Venugopal Rao,[2] M. Ghanashyam Krishna[1,3,*]

[1]School of Physics, [2]Advanced Centre of Research in High Energy Materials (ACRHEM),
[3]Centre for Advanced Studies in Electronic Sciences and Technology (CASEST),
University of Hyderabad, Hyderabad 500046, Telangana, India.
*Corresponding Author: email: mgksp@uohyd.ernet.in

**Abstract**

In this work, ZnO nanowire arrays decorated with titanium nitride (TiN) nanoparticles as surface enhanced Raman scattering (SERS) substrates is demonstrated. ZnO nanowires were grown by hydrothermal synthesis while ~100 nm TiN nanoparticles were obtained by grinding commercial powders for several hours. They were then decorated on the ZnO nanowire arrays using acetone as the medium. Scanning electron microscopy confirmed the presence of TiN nanoparticles on the ZnO nanowires. TiN nanoparticles exhibited localized surface plasmon resonances at 430, 520 and 600 nm. SERS experiments using Nile Blue and Methylene Blue as the analyte molecules showed significant enhancement in the Raman signals. It is shown that the origin of the SERS effect is chemical in nature, due to charge transfer between the analyte molecule and the TiN nanoparticles. The current work, thus, represents a simple, cost-effective and facile method for the fabrication of TiN based surface enhanced Raman scattering substrates.

**Keywords:** ZnO; TiN; Spectroscopy; optical properties; sensors

## 1. Introduction

Surface enhanced Raman scattering (SERS) and Surface enhanced resonance Raman scattering (SERRS) have emerged as important tools for the detection of chemical and biological molecules. Enhancement of the Raman signals is generally achieved by decorating surfaces with Cu, Au or Ag nanoparticles. It has been observed that most of the SERS enhancement occurs in the top layer of the metal layers. Hence, the approach is to employ metal nanoparticle decorated surfaces where in plasmon resonances can be tuned by varying the shape, size and inter-particle distance [1-11]. Recent work has shown that titanium nitride (TiN) thin films display a strong absorption peak related to plasmonic resonance at, approximately 530 nm which is very close to that of Au nanostructures [12,13]. In addition, the fact that, TiN possesses electrical conductivity which is almost the same as that of Au makes it an attractive alternative for many applications. As a result, TiN thin films and nanostructures can, indeed, be used as SERS substrates. This has been demonstrated by the ability to detect molecules such as R6G [14-21]. However, complex processing techniques such as sputter deposition are required to produce the reported materials. It is quite well documented that TiN has several sub-stoichiometric compositions and the growth of the stoichiometric form requires precise process control [22]. Therefore, although previous reports show the promise of TiN for SERS applications, they also point to the need for different strategies to expand the applicability. Hence, in this work TiN nanoparticle decorated ZnO nanowires are investigated for application as SERS substrates. To the best of the author's knowledge there are no previous reports on this subject.

## 2. Experimental

The ZnO nanowires are prepared by hydrothermal synthesis on ZnO thin film coated glass substrates, as described earlier [23]. Briefly, 100 nm thickness ZnO thin films are deposited on glass slides followed by annealing in air at 400°C for 2 hrs. ZnO nanowires were grown on these films by hydrothermal synthesis at 120°C for 3hrs in a 100 mL Teflon-lined stainless-steel autoclave. The precursors are aqueous solution of 25 mM each of Zinc acetate dehydrate [$Zn(O_2CCH_3)_2(H_2O)_2$, 99.0%- ZnAcD] and 25 mM hexamethylenetetramine (HMTA) [$(CH_2)_6N_4$, 99.0 %]. TiN nanoparticles (NPs) were prepared by hand-milling commercially purchased powders that contained particles of 3-5µm in diameter. After several hours of milling the particle size of the powders was reduced to 30-100 nm. These nanoparticles were then decorated on the ZnO nanowires using acetone as the medium. Field emission scanning electron microscopy (FE-SEM) images were acquired in a Model Ultra-55 microscope of Carl Zeiss. X-ray diffraction patterns were recorded in a PANalytical powder x-ray diffractometer using Cu K$\alpha$ radiation of wavelength =0.15408 nm. UV-visible-NIR spectra were recorded in a JASCO (Model V-670) spectrophotometer operating in the wavelength range of 190-2500nm. Raman and SERS measurements were performed with a LabRAM -Horiba Jobin Yvon spectrometer. A wavelength of 532 nm was used while the light was focused via a 50 X, 0.5 NA microscope objective. The SERS spectra of various analyte molecules (MB and NB) were collected and each Raman spectrum is a result of 5 accumulations with an acquisition time of 5s. The SERS substrates were prepared by simple drop casting 10 μL of MB/NB solution on TiN-ZnO samples.

## 3. Results and Discussion

The top view of a typical ZnO nanowire surface shown in fig. 1(a) indicates that the nanowires are of diameter between 30-100 nm. The nanowires are fairly densely packed and are aligned at different angles. Cross-sectional microscopy analysis reported earlier showed that the nanowires are reasonably vertically aligned [23]. The as-received TiN powders were subjected grinding for several hours. Many of the particles which were initially of the order 3-5μm were reduced to values of the order of 100 nm or less after grinding, as displayed in fig. 1(b). The nanoparticles were dispersed on the surface of ZnO nanowires using acetone as the medium, followed by drying in air. Closer examination of the particles, in fig.1(c) indicates that there is agglomeration of the particles at some locations and the shapes are non-spherical. The ZnO nanowires underneath the nanoparticles are clearly visible in these images. This method is, thus, a facile and simple technique to produce TiN nanoparticles and decorate them on ZnO nanowire surfaces. The method also eliminates the need for using complex processes to produce nanoparticles.

The X-ray diffraction pattern of the pristine TiN nanoparticles is shown in fig. 2(a). It is evident that the nanoparticles crystallize in the FCC structure with crystallite size of the order of 35 nm. The highest intensity reflection is from the {200} plane. When these nanoparticles are decorated on the ZnO nanowire surface, the XRD pattern is dominated by the peaks from ZnO nanowires. However, a low intensity peak from the {200} plane of FCC TiN is still visible. It can be inferred from the XRD pattern that no chemical reaction between the TiN nanoparticles fig. 2(b) for the TiN nanoparticle decorated ZnO nanowire surface. Constable et al. [24] observed peaks at 215, 327, 566, and 609 cm$^{-1}$ in the Raman spectra of TiN films and these modes were assigned to the transverse acoustic (TA), longitudinal acoustic (LA), second-order acoustic (2A), and transverse optical (TO) modes, respectively. In a report by Subramanian and Jayachandran [25], these features were observed at 320, 440 and 570 cm$^{-1}$, respectively. Thus, in the present case the peaks at 330, 559 and 662 cm$^{-1}$ are attributed to TiN nanoparticles while the peaks 374, 436 and 478 cm$^{-1}$ are attributed to ZnO. The most intense peak observed at 436 cm$^{-1}$ is the $E_2$ high mode of wurtzite ZnO. The other peaks at higher wavenumbers may have contributions from both materials. It is evident from the data presented so far, that TiN nanoparticles are formed on the surface of the ZnO nanowires and there is no chemical reaction between the two compounds. The optical absorption spectrum of TiN nanoparticles dispersed on a glass slide in the form of a thin film were recorded on a UV-Vis-Near IR spectrophotometer and shown in fig. 2(c). The spectrum displays a broad localized surface plasmon resonance (LSPR) at 430 nm and low intensity resonances at 520 and 600 nm. There is a blue shift in the most intense LSPR peak from the reported values [12,13] which can be attributed to the relatively small size of the nanoparticles. The multiple resonances are attributed to the non-spherical shape of the particles. Two molecules, Nile Blue (NB) and Methylene Blue (MB), were selected as prototypes to investigate the efficacy of the ZnO/TiN structures for SERS applications. The Raman spectrum of NB, recorded by dispersing it in the form of a film on a glass slide and displayed in fig. 3(a), consisted of low intensity modes at 382, 472, 553 and 653 cm$^{-1}$, and all of them matched with previous reports. However, the most significant feature occurs at 592 cm$^{-1}$, which is assigned to C-C-C and C-N-C deformations [26]. In contrast, the Raman spectrum of MB in fig. 3(b) exhibits several low intensity features at 450, 664, 716, 1038, 1154, 1390, 1430 and 1500 cm$^{-1}$, but the signature peak occurs at 1622 cm$^{-1}$ which is assigned to ring stretching of C-C mode, as shown in in fig. 3(b). All the characteristic peak were well matched with previous reports [27]. Thus, when these spectra are compared with the Raman spectra of TiN nanoparticle decorated ZnO nanowires presented in fig. 2(b), it is evident that the signature peaks of both molecules (NB and MB) are clearly distinct from that of ZnO and TiN. As a consequence, the SERS effect can be unambiguously studied.

The control experiment to investigate the SERS effect was carried out directly on ZnO nanowires (ZNW) at a concentration of 5 μM NB and MB. The spectra displayed in fig.4, do not show any evidence for enhancement of Raman signatures of either the NB or MB molecule. The only peak is at 436 cm$^{-1}$ corresponding to wurtzite ZnO. Evidently, the pristine ZNW surface is not suitable as a SERS substrate.

The SERS spectra for different concentrations of NB on TiN nanoparticles decorated ZnO nanowires (termed as T-ZNW) is shown fig. 5. It is observed that at a high concentration of 5μM, there is decrease in the intensity of the ZnO peak at 438 cm$^{-1}$ and this is accompanied by an increase in the intensity of the NB signature at 592 cm$^{-1}$, compared to the intensities on ZNW. At a NB concentration of 1 μM, on T-ZNW, there is complete quenching of the Raman modes relating to it and only one, high intensity peak at 438 cm$^{-1}$, attributed to wurtzite ZnO is visible. The Raman spectrum of 0.5mM NB dispersed on glass slide, presented earlier in fig. 3 is also displayed for comparison.

The SERS spectra of MB adsorbed on T-ZNW at different concentrations is shown in fig.6. At 5 μM concentration of MB, the intensity of this feature at 1622 cm$^{-1}$ is significantly increased. Further there is decrease in the intensity of the wurtzite ZnO mode, which now occurs at 443 cm$^{-1}$, as compared to the intensity on ZNW shown in fig.4. Further decrease in MB concentration to 1 μM, results in suppression of its Raman modes and only the wurtzite ZnO peak at 443cm$^{-1}$ is visible. In both cases, there is enhancement at 5 μM concentration while the Raman signals of the analyte molecules are quenched at 1 μM concentration. It is also evident that even at a high concentration of 0.5 mM, there is no enhancement of the Raman signals when the analyte molecules are adsorbed directly on the ZnO nanowires. Another interesting aspect is the shift of the most intense peak of ZnO to higher wavenumbers.

The enhancement factors were calculated at a concentration of 5 μM for both molecules using the expression [1, 2]

$$E_f = \frac{I_{SERS}}{I_{Raman}} \times \frac{C_{Raman}}{C_{SERS}} \qquad (1)$$

where $E_f$ is the enhancement factor, $I_{SERS}$ is the SERS band intensity of probe molecules [MB (1622 cm$^{-1}$) and NB (592 cm$^{-1}$)] using the synthesized TiN-ZnO substrate, $I_{Raman}$ is the Raman intensity of an probe molecule on a glass slide (without using substrate), $C_{SERS}$ represents the corresponding concentration of an probe molecule on a TiN-ZnO substrate (10$^{-6}$ M) and $C_{Raman}$ is the concentration of an probe molecule on a glass slide (10$^{-3}$ M, without substrate), which produces the Raman signal, $I_{Raman}$.

The EF values obtained are 94 for NB (592 cm$^{-1}$) and 28 for MB (1622 cm$^{-1}$) at 5μM. It is well established that there are two main mechanisms of surface enhancement of Raman signals [1-11]. They are either electromagnetic or chemical in nature. The electromagnetic enhancement of signals involves the coupling of the electric fields related to surface plasmons of the nanoparticles with incident radiation. The strength of coupling varies as fourth power of the local electric field. As a result, very small variation in the local field can cause very significant changes in the SERS signals. Chemical enhancement is, in contrast, a weaker effect relying on formation of complexes on the nanoparticle surface, charge transfer or charge transfer resonances induced by molecular adsorption. Due to the differences in their origins, the enhancement factors observed due to the EM effect are large in magnitude ($10^4$ or higher) whereas the chemical effect enhancement factors are of the order of $10^3$, at maximum. It would, thus, appear that in the present case the SERS effect is more chemical in nature. Interestingly, Wei et al. [18] have observed a combination of the electromagnetic and charge transfer effect in the TiN thin film based SERS substrates that were used to detect R6G.

TiN has been shown to be a good catalyst for photodegradation of molecules such as methylene blue [28]. The essential requirement is the adsorption of the molecule to be photodegraded onto the catalyst material. This provides some insight into the reason for SERS effect being chemical in nature in the present case. In photocatalysis, the molecule absorbs the incident radiation to generate electron-hole pairs which is followed by separation of the excited charges and charge transfer to the surface of the photocatalyst. Finally, these charges are used for redox reactions [29]. As stated earlier, chemical enhancement of SERS signals is charge transfer between the adsorbed analyte molecule and the plasmonic nanoparticle. It would, thus, be, reasonable to assume that this process is occurring in the present case. As the concentration of the dye molecules on the T-ZNW surface is decreased the intensity of their signature peak also decreases, probably due to reduced adsorption and, therefore, weaker charge transfer. Nile blue has the chemical formula, $C_{20}H_{20}ClN_3O$ and belongs to the family of phenoxazine and benzo-phenoxazine based dyes. Methylene blue has the chemical formula $C_{16}H_{18}ClN_3S$ and is also known as methylthioninium chloride. Since in both compounds the chloride ions are present, it is hypothesized that the affinity of TiN to chloride ions might cause the binding of the dye molecules to the nanoparticles leading to the chemical SERS effect.

Though the enhancements are moderate in the present case it is highly possible to improve these values. The proof-of-concept study indicates that further enhancements will be possible with these substrates by optimizing the size, shape and inter-particle distance of the TiN nanoparticles. The main advantages of these substrates are the ease of preparing large-area substrates and the low costs- involved. Further detailed studies with different analyte molecules (including explosives and pesticides) along with reproducibility and recyclability data will bring out the exact capability of these substrates to the fore.

## 4. Conclusions

TiN nanoparticle decorated ZnO nanowires are synthesized by a simple process. The TiN nanoparticles are 30-100 nm in size and exhibit localized surface plasmon resonances at wavelengths between 400-600 nm. The application of these substrates as surface enhanced Raman scattering substrates is demonstrated by detecting nile blue and methylene blue molecules on the nanoparticles. It is demonstrated that the origin of SERS in this system is chemical in nature. The facile approach to synthesis indicates promise for cost effective scale-up.

**Acknowledgements**

Y.Rajesh acknowledges a NFHE Ph.D. fellowship awarded by UGC. M.S.S. Bharati acknowledges DRDO financial support through ACRHEM, University of Hyderabad. Center for Nanotechnology and School of Physics, University of Hyderabad, India are acknowledged for facilities. The support of DST-PURSE, UGC-DRS, UGC-NRC programmes is also acknowledged.

**References**

[1] J. Langer, D. Jimenez de Aberasturi, J. Aizpurua, R.A. Alvarez-Puebla, B. Auguié, J.J. Baumberg, G.C. Bazan, S.E.Bell, A. Boisen, A.G.Brolo, J. Choo *et al.*, "Present and future of surface-enhanced Raman scattering." *ACS Nano* 14 (2019) 28-117.

[2] A.I. Pérez-Jiménez, D. Lyu, Z. Lu, G. Liu, and B. Ren, "Surface-enhanced Raman spectroscopy: benefits, trade-offs and future developments." *Chem.Sci.* 11.18 (2020) 4563-4577.

[3] J. Kim, Y. Jang, N.J. Kim, H. Kim, G.C. Yi, Y. Shin, M.H. Kim, and S. Yoon, "Study of chemical enhancement mechanism in nonplasmonic surface enhanced Raman spectroscopy (SERS)." *Front. Chem.* 7 (2019) 582.

[4] E.C. Le Ru, S.A. Meyer, C. Artur, P.G. Etchegoin, J. Grand, P. Lang, F. Maurel, "Experimental demonstration of surface selection rules for SERS on flat metallic surfaces." *Chem. Commun.* 47 (2011) 3903-3905.

[5] E.C. Le Ru, P.G. Etchegoin. "Quantifying SERS enhancements." *MRS Bull.* 38 (2013) 631.

[6] G. McNay, D. Eustace, W.E. Smith, K. Faulds, D. Graham, "Surface-enhanced Raman scattering (SERS) and surface-enhanced resonance Raman scattering (SERRS): a review of applications." *Appl. Spectrosc.* 65 (2011) 825-837.

[7] Y. Liu, H. Ma, X. X. Han, B. Zhao, "Metal-semiconductor heterostructures for surface-enhanced Raman scattering: synergistic contribution of plasmon and charge Transfer" *J.Mater. Horiz.*, 2020. https://doi.org/10.1039/D0MH01356K

[8] Y. Wu, M. Yang, W. U. Tyler, A. M. Martín, C. Zhu, G. C. Schatz, R. P. Van Duyne, "SERS Study of the Mechanism of Plasmon-Driven Hot Electron Transfer between Gold Nanoparticles and PCBM." *J.Phys.Chem.C,* 123 *(*2019)**,** 29908-29915.

[9] S.-Y. Ding, J. Yi, J.-F. Li, B. Ren, D.-Y. Wu, R. Panneerselvam and Z.-Q. Tian, "Nanostructure-based plasmon-enhanced Raman spectroscopy for surface analysis of materials," *Nat. Rev. Mater.*, 1 (2016), 16021.

[10] S. Hamad, G.K. Podagatlapalli, M.A. Mohiddon, S. Venugopal Rao, "Surface enhanced fluorescence from corroles and SERS studies of explosives using copper nanostructures." *Chem. Phys. Lett.* 621 (2015) 171-176.

[11] U.P. Shaik, S. Hamad, M. Ahamad Mohiddon, S. Venugopal Rao, M. Ghanashyam Krishna, "Morphologically manipulated Ag/ZnO nanostructures as surface enhanced Raman scattering probes for explosives detection." *J.Appl.Phys.* 119 (2016) 093103.

[12] G. V. Naik, J. L. Schroeder, X. Ni, A. V. Kildishev, T. D. Sands, A. Boltasseva, "Titanium nitride as a plasmonic material for visible and near-infrared wavelengths." *Opt. Mater. Exp.*, 2(2012), 478-489.

[13] M. Gioti, J. Arvanitidis, D. Christofilos, K. Chaudhuri, T. Zorba, G. Abadias, D. Gall, V.M. Shalaev, A. Boltasseva, P. Patsalas, "Plasmonic and phononic properties of epitaxial conductive transition metal nitrides," *J. Opt.* 22 (2020) 084001.

[14] T. Fu, Y. Chen, C. Du, W. Yang, R. Zhang, L. Sun, D. Shi, "Numerical investigation of plasmon sensitivity and surface-enhanced Raman scattering enhancement of individual TiN nanosphere multimers." *Nanotechnology* 31(2020) 135210.

[15] Y. Ma, D. Sikdar, A. Fedosyuk, L. Velleman, D.J. Klemme, S.H. Oh, A.R. Kucernak, A.A. Kornyshev, J.B. Edel, "Electrotunable Nanoplasmonics for Amplified Surface Enhanced Raman Spectroscopy." *ACS Nano* 14 (2019) 328-336.

[16] F. Zhao, X. Xue, W. Fu, Y. Liu, Y. Ling, Z. Zhang, "TiN Nanorods as Effective Substrate for Surface-Enhanced Raman Scattering." *J. Phys.Chem. C* 123 (2019) 29353-29359.

[17] A.A. Popov, G. Tselikov, N. Dumas, C. Berard, K. Metwally, N. Jones, A. Al-Kattan, B. Larrat, D. Braguer, S. Mensah, A. Da Silva, "Laser-synthesized TiN nanoparticles as promising plasmonic alternative for biomedical applications." *Sci.Rep.* 9 (2019) 1-11.

[18] H. Wei, M. Wu, Z. Dong, Y. Chen, J. Bu, J. Lin, Y. Yu, Y. Wei, Y. Cui, R. Wang, "Composition, microstructure and SERS properties of titanium nitride thin film prepared via nitridation of sol–gel derived titania thin films." *J. Raman Spectrosc.* 48 (2017) 578-585.

[19] I. Lorite, A. Serrano, A. Schwartzberg, J. Bueno and J.L. Costa-Krämer, "Surface enhanced Raman spectroscopy by titanium nitride non-continuous thin films." *Thin Solid Films* 531 (2013) 144-146.

[20] J. Zhao, J. Lin, H. Wei, X. Li, W. Zhang, G. Zhao, J. Bu, Y. Chen, "Surface enhanced Raman scattering substrates based on titanium nitride nanorods." *Opt.Mater.* 47 (2015) 219-224.

[21] N. Kaisar, Y.T. Huang, S. Jou, H.F. Kuo, B.R. Huang, C.C. Chen, Y.F. Hsieh, Y.C. Chung, "Surface-enhanced Raman scattering substrates of flat and wrinkly titanium nitride thin films by sputter deposition." *Surf. Coat. Technol.* 337 (2018) 434-438.

[22] K. Vasu, M. Ghanashyam Krishna, K.A. Padmanabhan, "Substrate-temperature dependent structure and composition variations in RF magnetron sputtered titanium nitride thin films." *Appl. Surf. Sci.* 257 (2011) 3069-3074.

[23] Y. Rajesh, S. K. Padhi, M. Ghanashyam Krishna, "ZnO thin film-nanowire array homo-structures with tunable photoluminescence and optical band gap." *RSC Adv.* 10 (2020) 25721-25729.

[24] C. P. Constable, J. Yarwood, W-D. Münz, "Raman microscopic studies of PVD hard coatings." *Surf. Coat. Technol.* 116 (1999) 155-159.

[25] B. Subramanian, M. Jayachandran "Characterization of reactive magnetron sputtered nanocrystalline titanium nitride (TiN) thin films with brush plated Ni interlayer." *J. Appl. Electrochem.* 37 (2007) 1069-1075.

[26] M.S.S. Bharati, C. Byram, S. Venugopal Rao, "Gold-nanoparticle-and nanostar-loaded paper-based SERS substrates for sensing nanogram-level Picric acid with a portable Raman spectrometer." *Bull. Mater. Sci.* 43.1 (2020) 53.

[27] M.S.S. Bharati, C. Byram, S.Venugopal Rao, "Explosives sensing using Ag–Cu alloy nanoparticles synthesized by femtosecond laser ablation and irradiation." *RSC Adv.* 9 (2019) 1517-1525.

[28] J.B. Yoo, H.J. Yoo, H.J. Jung, H.S. Kim, S. Bang, J. Choi, H. Suh, J.H. Lee, J.G. Kim, and N.H. Hur, "Titanium oxynitride microspheres with the rock-salt structure for use as visible-light photocatalysts." *J. Mater. Chem.A* 4 (2016) 869-876.

[29] S. Zhu, D. Wang "Photocatalysis: basic principles, diverse forms of implementations and emerging scientific opportunities." *Adv.Energy Mater.* 7 (2017) 1700841.

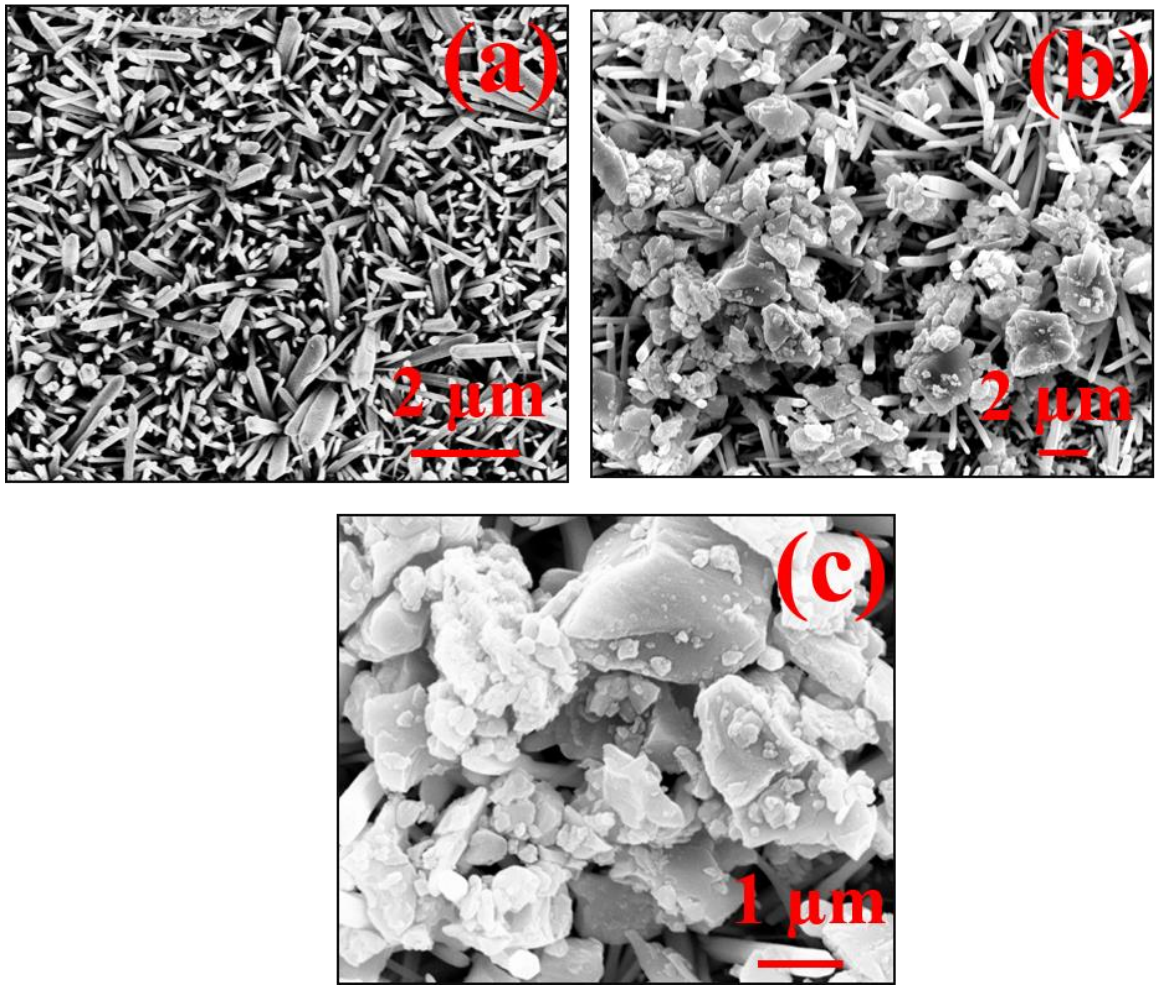

**Figure 1** FE-SEM images of (a) ZnO nanowires and (b) and (c) TiN nanoparticles decorated ZnO nanowires at different magnifications.

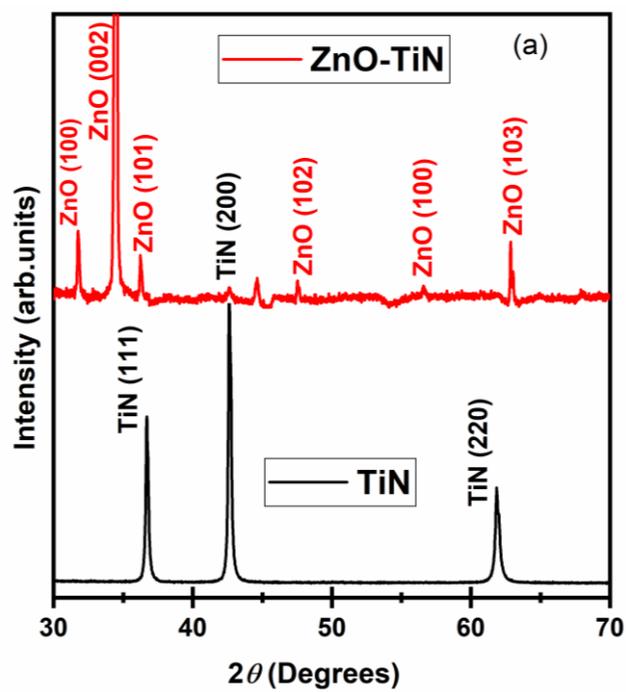

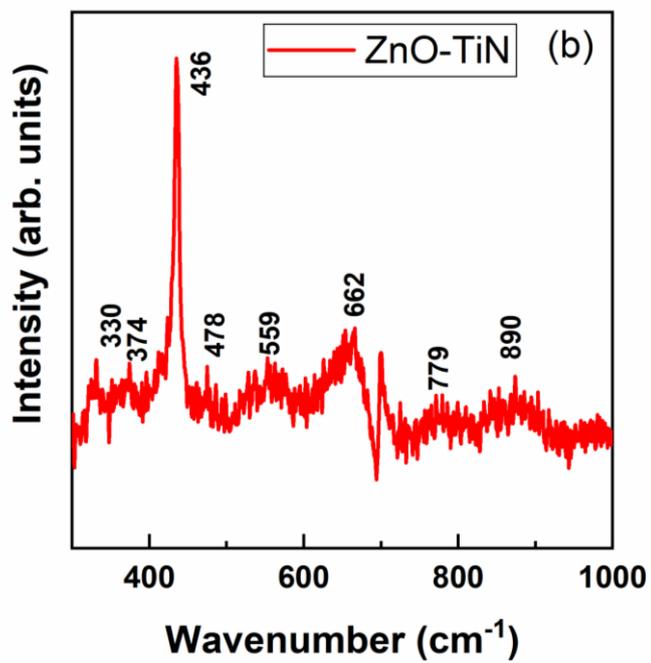

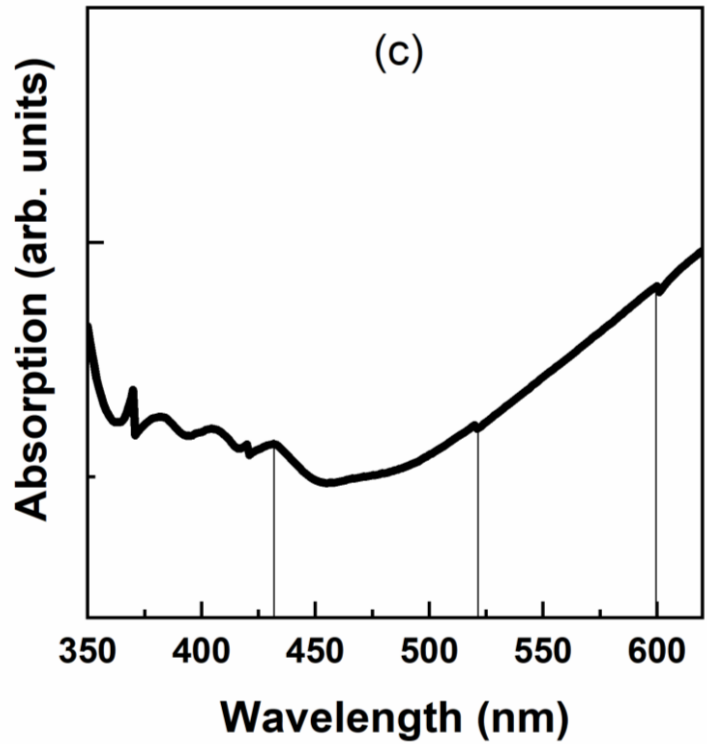

**Figure 2** (a) XRD patterns of TiN nanoparticles and ZnO nanowires decorated with TiN nanoparticles, (b) Raman spectrum of TiN nanoparticles decorated with ZnO nanowires and (c) optical absorption spectrum of TiN nanoparticles

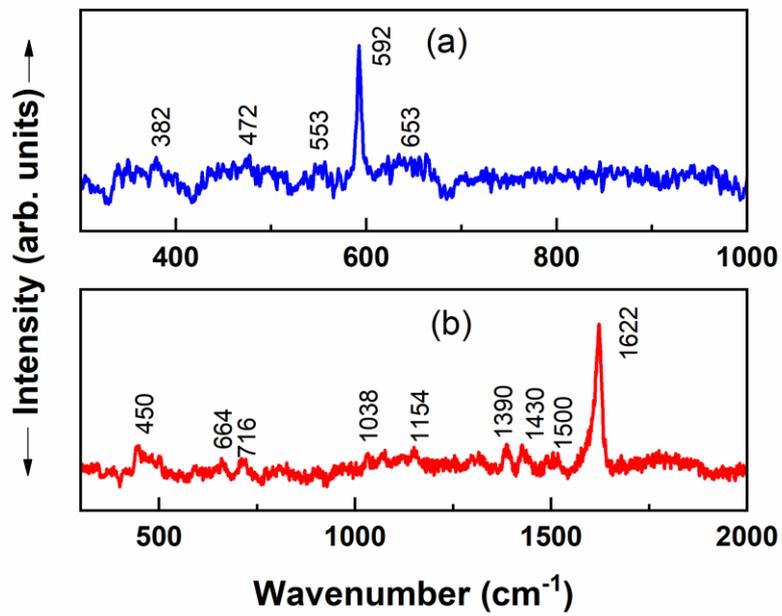

**Figure 3** Raman spectra of (a) Nile blue and (b) Methylene blue on glass slides.

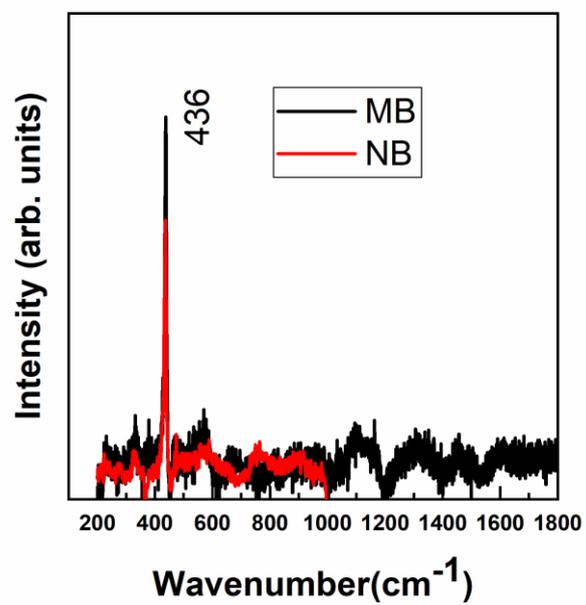

**Figure 4** Raman spectra of Nile blue (NB) and Methylene Blue (MB) adsorbed directly on ZnO nanowires at 5 μM concentration

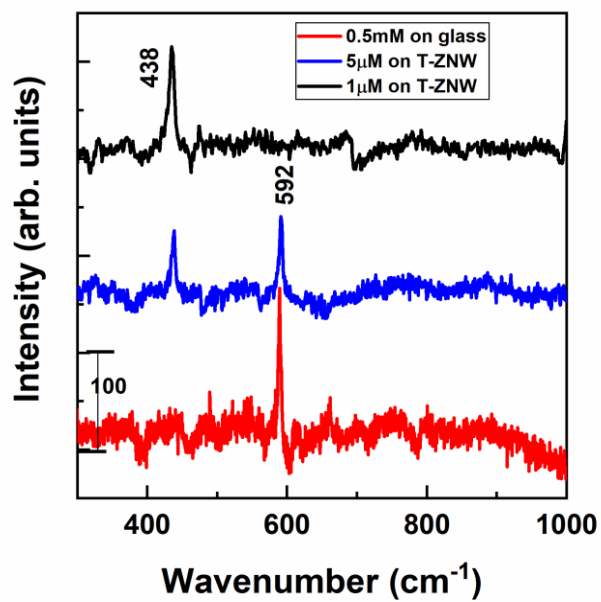

**Figure 5** SERS spectra of Nile Blue adsorbed on TiN nanoparticles decorated ZnO nanowires. The red curve corresponds to the Raman spectrum of 0.5 mM NB dispersed directly on glass, for comparison

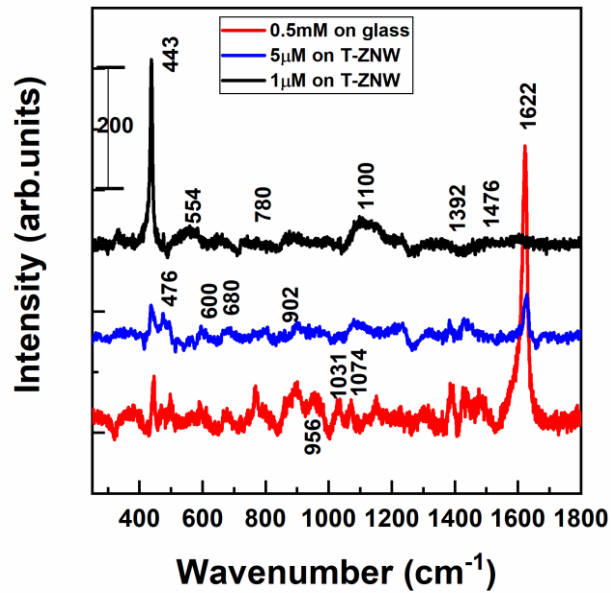

**Figure 6** SERS spectra of Methylene Blue adsorbed on TiN nanoparticles decorated ZnO nanowires. The red curve corresponds to the Raman spectrum of 0.5mM MB dispersed directly on glass, for comparison.